\documentclass[aps,prl,twocolumn,superscriptaddress,showpacs]{revtex4-1}

\usepackage{graphicx}
\usepackage{dcolumn}
\usepackage{bm}
\usepackage{cancel}
\usepackage{color}
\usepackage{enumitem}

\setlist[enumerate]{nosep}

\newcommand{\bfa}{BaFe$_2$As$_2$}

\newcommand{\AFS}{A$_{x}$Fe$_{2-y}$Se$_2$}

\newcommand{\sto}{SrTiO$_3$}

\newcommand{\ef}{$E_F$}
\newcommand{\kf}{$k_F$}
\newcommand{\kz}{$k_z$}

\newcommand{\tc}{$T_c$}
\newcommand{\ts}{$T_S$}

\newcommand{\cfour}{C$_4$}
\newcommand{\ctwo}{C$_2$}
\newcommand{\dxy}{$d_{xy}$}
\newcommand{\dxz}{$d_{xz}$}
\newcommand{\dyz}{$d_{yz}$}

\newcommand{\g}{$\Gamma$}

\newcommand{\gm}{$\Gamma-M$}
\newcommand{\gmx}{$\Gamma-M_X$}
\newcommand{\gmy}{$\Gamma-M_Y$}
\newcommand{\mxgtwo}{$M_X-\Gamma_2$}
\newcommand{\mygtwo}{$M_Y-\Gamma_2$}
\newcommand{\mx}{$M_X$}
\newcommand{\my}{$M_Y$}

\begin{document}

\title{The Nematic Energy Scale and the Missing Electron Pocket in FeSe}

\author{M. Yi}
\email{mingyi@rice.edu}
\affiliation{Department of Physics and Astronomy, Rice University, Houston, TX 77005, USA}
\affiliation{Department of Physics, University of California Berkeley, Berkeley, CA 94720, USA}
\author{H. Pfau}
\affiliation{Advanced Light Source, Lawrence Berkeley National Laboratory, Berkeley, Berkeley, CA 94720, USA}
\affiliation{Stanford Institute of Materials and Energy Sciences, SLAC National Accelerator Laboratory, Menlo Park, CA 94025, USA}
\affiliation{Departments of Physics and Applied Physics, and Geballe Laboratory for Advanced Materials, Stanford University, Stanford, CA 94305, USA}
\author{Y. Zhang}
\affiliation{International Center for Quantum Materials, School of Physics, Peking University, Beijing 100871, China}
\affiliation{Collaborative Innovation Center of Quantum Matter, Beijing 100871, China}
\author{Y. He}
\affiliation{Stanford Institute of Materials and Energy Sciences, SLAC National Accelerator Laboratory, Menlo Park, CA 94025, USA}
\affiliation{Department of Physics, University of California Berkeley, Berkeley, CA 94720, USA}
\author{H. Wu}
\affiliation{Department of Physics and Astronomy, Rice University, Houston, TX 77005, USA}
\author{T. Chen}
\affiliation{Department of Physics and Astronomy, Rice University, Houston, TX 77005, USA}
\author{Z. R. Ye}
\affiliation{International Center for Quantum Materials, School of Physics, Peking University, Beijing 100871, China}
\author{M. Hashimoto}
\affiliation{Stanford Synchrotron Radiation Lightsource, SLAC National Accelerator Laboratory, Menlo Park, CA 94025, USA}
\author{R. Yu}
\affiliation{Department of Physics, Renmin University of China, Beijing 100872, China}
\author{Q. Si}
\affiliation{Department of Physics and Astronomy, Rice University, Houston, TX 77005, USA}
\author{D.-H. Lee}
\affiliation{Department of Physics, University of California Berkeley, Berkeley, CA 94720, USA}
\affiliation{Materials Sciences Division, Lawrence Berkeley National Laboratory, Berkeley, CA 94720, USA}
\author{Pengcheng Dai}
\affiliation{Department of Physics and Astronomy, Rice University, Houston, TX 77005, USA}
\author{Z.-X. Shen}
\affiliation{Stanford Institute of Materials and Energy Sciences, SLAC National Accelerator Laboratory, Menlo Park, CA 94025, USA}
\affiliation{Departments of Physics and Applied Physics, and Geballe Laboratory for Advanced Materials, Stanford University, Stanford, CA 94305, USA}
\author{D. H. Lu}
\email{dhlu@slac.stanford.edu}
\affiliation{Stanford Synchrotron Radiation Lightsource, SLAC National Accelerator Laboratory, Menlo Park, CA 94025, USA}
\author{R. J. Birgeneau}
\email{robertjb@berkeley.edu}
\affiliation{Department of Physics, University of California Berkeley, Berkeley, CA 94720, USA}
\affiliation{Materials Sciences Division, Lawrence Berkeley National Laboratory, Berkeley, CA 94720, USA}
\affiliation{Department of Materials Science and Engineering, University of California, Berkeley, CA 94720, USA}

\date{\today}

\begin{abstract}
%
Superconductivity emerges in proximity to a nematic phase in most iron-based superconductors. It is therefore important to understand the impact of nematicity on the electronic structure. Orbital assignment and tracking across the nematic phase transition proved to be challenging due to the multiband nature of iron-based superconductors and twinning effects. Here we report a detailed study of the electronic structure of fully detwnned FeSe across the nematic phase transition using angle-resolved photoemission spectroscopy. We clearly observe a nematicity-driven band-reconstruction involving \dxz, \dyz~and \dxy~orbitals. The nematic energy scale between \dxz~and \dyz~bands reach a maximum of 50meV at the Brillouin zone corner. We are also able to track the \dxz~electron pocket across the nematic transition and explain its absence in the nematic state. Our comprehensive data of the electronic structure provide an accurate basis for theoretical models of the superconducting pairing in FeSe.
\end{abstract}


\maketitle

\section{Introduction}

Electronic nematicity, defined as the breaking of the four-fold rotational symmetry by the electronic degree of freedom, has been widely found in iron-based superconductor (FeSC) families~\cite{Johnston2010,Paglione2010,Fernandes2014d}. Its experimental manifestations in FeSCs include a tetragonal-to-orthorhombic structural transition~\cite{Johnston2010}, rotational symmetry breaking detected by probes sensitive to the charge and orbital degrees of freedom~\cite{Chu2010,Chu2012,Dusza2011,Yi2011,Chuang2010}, and anisotropy in the spin susceptibility~\cite{Kasahara2012,Fu2012,Lu2014c}. The electronic origin of the nematicity is demonstrated by a divergent susceptibility of the resistivity anisotropy~\cite{Chu2012}. In almost all FeSCs, the nematic order is strongly coupled to a collinear antiferromagnetic order onsetting simultaneously or slightly below the structural transition~\cite{Dai2015}. This strong coupling between the spin, orbital, and lattice degrees of freedom has led to an intense debate on the driving mechanism~\cite{Fernandes2014d}, with proposals based on orbital order~\cite{Lee2009,Chen2010,Kontani2011a} or spin nematicity~\cite{Fang2008,Xu2008a,Dai2009,Lv2010,Fernandes2012a}. An exception to this strong coupling of the nematic order and magnetic order is iron selenium (FeSe). Structurally the simplest FeSC, FeSe is the only compound that exhibits a nematic order (\ts~= $\sim$90 K)~\cite{Margadonna2008} without a long range magnetic order. Therefore FeSe provides a unique opportunity to explore the effect of nematicity disentangled from that of the static magnetic order. 

Furthermore, FeSe also provides a platform to study the interaction of nematicity with superconductivity. Bulk FeSe exhibits superconductivity below \tc~= 8 K~\cite{Hsu2008}, and is highly tunable. Under hydrostatic pressure, the modest bulk \tc~can rise up to 37 K~\cite{Medvedev2009c}. Intercalation with atoms or molecules between the FeSe layers that introduce electron doping such as in bulk \AFS~(A = K, Rb, Cs)~\cite{Guo2010,Dagotto2013} and (Li$_{1-x}$Fe$_x$)OHFeSe~\cite{Lu2014d,Dong2015c} can enhance the \tc~up to 40 K~\cite{Lu2014d}, as can tuning the surface charge carrier by surface-doping with alkaline metals~\cite{Ye2015,Wen2016a,Seo2016}. When grown as a monolayer film on \sto, FeSe is generally considered capable of superconductivity above 60 K~\cite{Wang2012l,Tan2013,He2013,Lee2014,Ge2015}. Interestingly, all of these methods suppress nematicity in the process.

It is therefore critical to understand the electronic structure of the nematic state in order to formulate a theoretical model for superconductivity in FeSe and other FeSCs. However, the multiorbital nature of FeSCs and twinning in the nematic state are significant challenges that prevented a unified description of the electronic structure in the nematic state. In particular, two aspects are actively debated in literature in the case of FeSe. i) The nematic energy scale of the orbital anisotropy between \dxz~and \dyz~is interpreted as either much larger than ($\sim$50 meV)~\cite{Nakayama2014a,Shimojima2014,Suzuki2015a,Zhang2015a,Watson2015,Xu2016b,Zhang2016c,Fanfarillo2016a,Coldea2017,Liu2018}, or on par with the lattice distortion and superconducting energy scales ($\leq$10 meV)~\cite{Fedorov2016,Watson2016c,Watson2017}. This discrepancy has caused a revisit of theoretical understanding of the electronic nematic order in FeSe. ii) One of the two electron Fermi pockets observed in the normal state has escaped detection entering the nematic phase~\cite{Watson2017}. While the cause of the missing electron pocket has remained elusive, the incorporation of its absence into theoretical models~\cite{Sprau2017,Hu2018,Yu2018,Kang2018} has been deemed necessary to reproduce the observed strongly anisotropic superconducting gap~\cite{Xu2016b,Sprau2017,Liu2018}.

To clarify these issues, we present high quality ARPES measurements of the electronic structure of completely detwinned FeSe. Our data unambiguously demonstrate a nematic energy splitting of 50 meV at the BZ corner between the \dxz~and \dyz~orbitals, consistent with other FeSCs where the nematic order is strongly coupled to a magnetic order. We also clearly follow the "missing" electron pocket from the normal state well into the nematic state and observe its disappearance via shrinking. We explain this behavior by a band inversion that occurs between the \dxz~electron band and the \dxy~hole band at the BZ corner---the \my~point. This band inversion opens up a hybridization gap between the \dxz~and \dxy~bands such that the electron band containing \dxz~character in the normal state is pushed up in energy across the Fermi level (\ef). Both these band structure effects strongly reduce the presence of \dxz~states near the \my~point in the nematic phase, which could cause a suppression of the inter-pocket scattering along the (0, $\pi$) direction in the superconducting state. In addition, by comparing the measured versus calculated bandwidths of each orbital, we estimate the \dxz~and \dyz~orbitals to have similar correlation strengths while \dxy~is more strongly correlated. Taking all the observations together, we provide a self-consistent picture of the effect of nematicity on the low energy electronic states of FeSe. Our results provide the basis for future theoretical models of FeSe.

\begin{figure}
\includegraphics[width=0.45\textwidth]{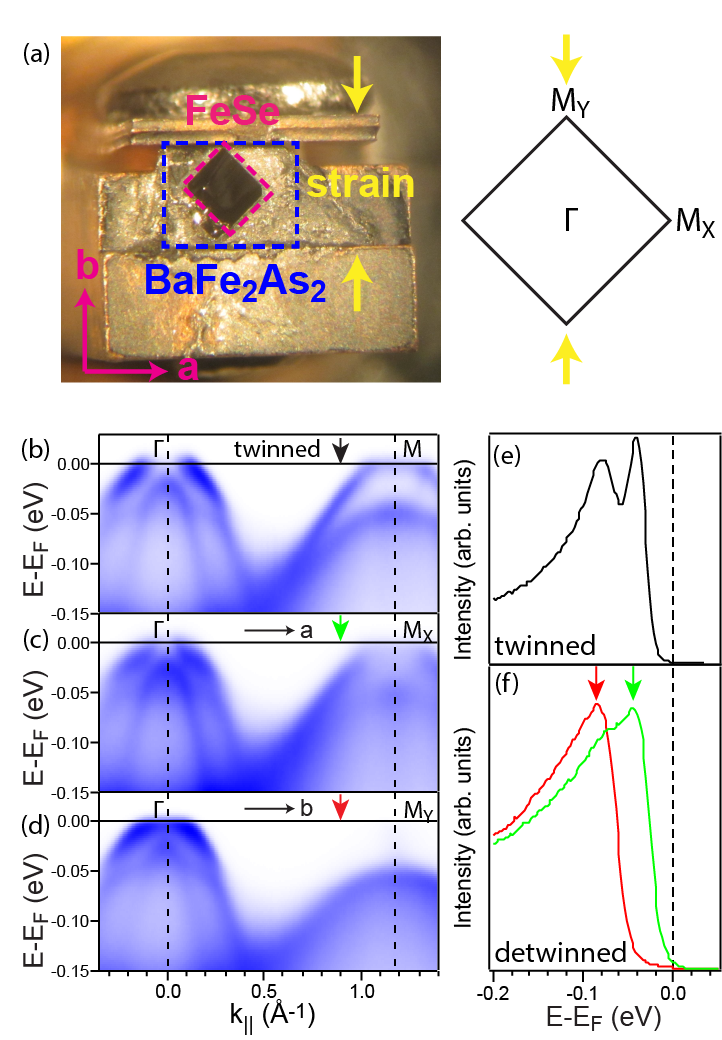}
\caption{\label{fig:fig1} Effectiveness of detwinning by uniaxial strain. (a) The mechanical uniaxial strain setup consists of a clamp that presses a single crystal of \bfa, which transfers the strain to the FeSe single crystal glued on top. Both the \bfa~and FeSe are oriented such that the Fe-Fe bond is aligned to the direction of strain. When cooled below \ts, the shorter (longer) Fe-Fe bond is along (perpendicular to) the strain direction, defining the \gmy~(\gmx) momentum direction. (b) ARPES spectra taken along the \gm~direction on a twinned FeSe. (c)-(d) ARPES spectra taken on detwinned FeSe along the \gmx~and \gmy~directions, respectively. (e)-(f) EDCs taken at the momentum pointed to by arrows in (b)-(d). All measurements were taken with 70 eV photons under odd polarization with respect to the cut direction.}
\end{figure}


\begin{figure*}
\includegraphics[width=\textwidth]{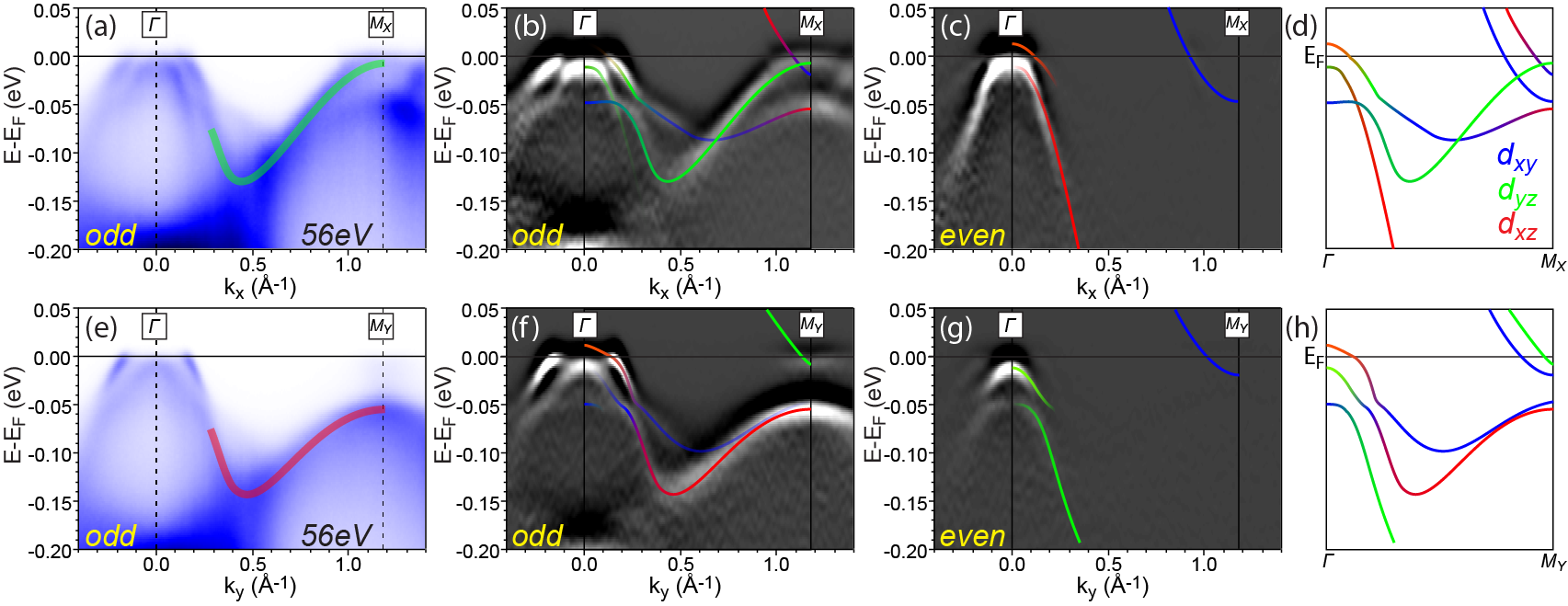}
\caption{\label{fig:fig2} Measured dispersions on detwinned FeSe at 56 eV (close to \kz~= $\pi$). (a) ARPES spectra measured along \gmx~with odd polarization. (b)-(c) Second energy derivatives of measured spectra along \gmx under odd and even polarization with respect to the cut direction, respectively. Schematic of bands of orbital symmetries with allowed intensity under each polarization is overlaid. (d) The complete band schematic from both polarizations is summarized for \gmx. (e)-(h) Similar measurement as (a)-(d) but for the \gmy~direction.}
\end{figure*}

\section{Methods}
High-quality single crystals of FeSe were grown by the chemical vapor transport method~\cite{Bohmer2016}. ARPES measurements were carried out at beamline 5-2 of the Stanford Synchrotron Radiation Lightsource using a SCIENTA D80 electron analyzer or a DA30 electron analyzer. The total energy resolution was set to 10 meV or better and the angular resolution was 0.1$^o$. Single crystals were cleaved \emph{in-situ} and measured at 15 K unless otherwise noted. All measurements were carried out in ultrahigh vacuum with a base pressure lower than 5x10$^{-11}$ torr. To detwin the FeSe crystals, we mount them in a mechanical detwin device~\cite{Yi2011} (Fig.~\ref{fig:fig1}a). We use single crystalline \bfa~as a substrate material to overcome the soft nature of FeSe. The tetragonal in-plane axes of both crystals are pre-aligned along the strain direction. Mechanical strain is then added to the \bfa~substrate, which is in turn transmitted to the FeSe crystal (Fig.~\ref{fig:fig1}a). Previous neutron diffraction experiments have shown that a single crystal of FeSe can be completely detwinned below the structural transition temperature of \bfa~\cite{Chen2019}. 

To demonstrate the effectiveness of this detwinning method, we compare the measured band dispersions along the orthogonal high symmetry directions \gmx~and \gmy~with that of a twinned sample (Fig.~\ref{fig:fig1}). The energy distribution curve (EDC) close to the M point of the twinned sample shows two peaks indicating the presence of two dominant bands (Fig.~\ref{fig:fig1}e). In contrast, each of the EDCs obtained on a detwinned sample along the two orthogonal momentum directions shows only one of the two peaks (Fig.~\ref{fig:fig1}f). The two bands therefore belong to two different domains. Importantly, for the EDC taken along \gmy~marked in red, there is no residual intensity shoulder at the energy where the peak from the other domain appears, indicating that the detwinning is complete. 

\section{Experimental Results}
\subsection{The nematic energy scale between \dxz/\dyz}

In the nematic state, the degeneracy of the \dxz~and \dyz~orbitals is lifted. The resulting band splitting between the \gmx~and \gmy~direction determines the energy scale of the nematic order. To determine this energy scale, we focus on the band dispersion of the \dxz~and \dyz~hole bands along  these two momentum directions shown in Fig.~\ref{fig:fig2}. We utilize selection rules for different light polarizations~\cite{Yi2011,Brouet2012} to identify the dominant orbital characters of bands observed on the detwinned crystal. Near the \g~point, three hole bands are resolved close to \ef, dominated by \dxz, \dyz, and \dxy~characters, as expected for all FeSCs~\cite{Yi2017}. Amongst these three, the \dxz~and \dyz~hole bands have comparable band velocities while the \dxy~hole band is much flatter and does not cross \ef. The \dxy~band tops at $\sim$50 meV below \ef~at \g, and crosses the \dxz~and \dyz~hole bands as it disperses towards the M point, leading to hybridization gaps. The identification of the orbital characters of these three hole bands near \g~agree amongst all previous ARPES reports on FeSe~\cite{Nakayama2014a,Shimojima2014,Suzuki2015a,Zhang2015a,Watson2015,Xu2016b,Watson2016c,Zhang2016c,Fanfarillo2016a,Watson2017,Coldea2017,Liu2018,Fedorov2016}, and is illustrated in a schematic in Fig.~\ref{fig:fig2}d,h. 

\begin{figure*}
\includegraphics[width=\textwidth]{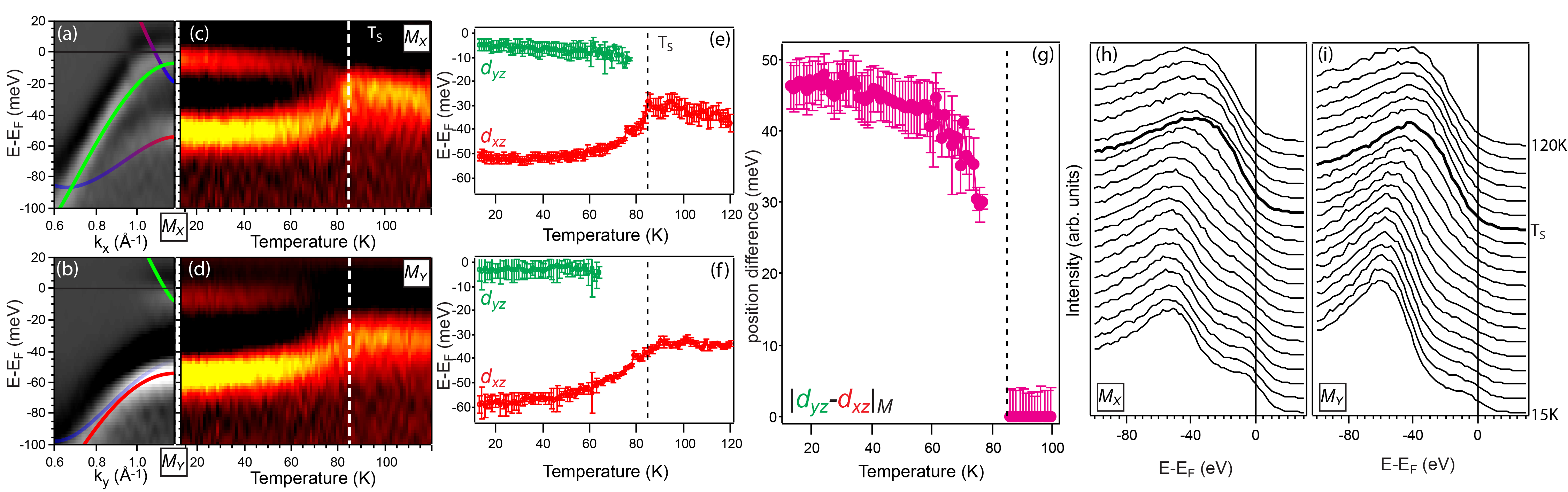}
\caption{\label{fig:fig3} Temperature evolution across the nematic transition on detwinned FeSe along \gmx~and \gmy. (a) Second energy derivative of spectra around the \mx~point along the $a$ direction, taken under odd polarization at 56 eV. (b) Corresponding measurement for the \my~point along the $b$ direction. (c) Temperature evolution of the EDC's second energy derivative taken at \mx. (d) Corresponding temperature evolution for the \my~point. (e) Fitted band positions for the \dyz~and \dxz~bands from (c). (f) Corresponding fits for (d). (g) The \dxz~and \dyz~band splitting as a function of temperature extracted for (e). (h)-(i) The raw EDCs at \mx~and \my~at selected temperatures.}
\end{figure*}

Following this identification towards the \mx~and \my~points, we see that the \dyz~hole band along \gmx~(marked green in Fig.~\ref{fig:fig2}a) deflects back up towards \ef~reaching a top near \ef~at the \mx~point. Similarly, along the orthogonal direction \gmy, the \dxz~hole band (marked red in Fig.~\ref{fig:fig2}e) deflects up in the same fashion, but topping at -50 meV at \my. This band assignment is also reinforced by polarization matrix element considerations. Under odd polarization with respect to \gmx, the \dyz~orbital has the strongest allowed intensity~\cite{Yi2011}. As the data along the \gmx~and \gmy~directions were recorded by rotating the crystal 90$^o$, by symmetry, under the same polarization, \dxz~orbital has equally strong intensity. In comparison, the \dxy~orbital in both measurements has much weaker expected intensity and cannot account for the strong hole like dispersion near the zone corner. This is consistent with the most intense spectral feature along \gmx~identified as \dyz~and the equally intense band reaching -50 meV at \my~identified as \dxz. The energy difference between the \dxz~and \dyz~bands at the M point is therefore 50 meV.

\subsection{Temperature evolution of electronic structure}
In this section we discuss in turn the following three observations of band structure evolution near the M point as temperature is lowered across the nematic phase transition on detwinned FeSe: 
\begin{enumerate}[label=\roman*]
\item Shifting down (up) of the \dxz~(\dyz) state at the M point
\item Expansion of the \dxy~portion of the electron pocket near the \mx~point
\item Shrinking of the electron pocket containing \dxz~orbital at the \my~point.
\end{enumerate}

First, to examine the behavior of the \dxz/\dyz~states at the M points, we present a detailed temperature-dependent measurement along the high symmetry directions \gmx~and \gmy~near the BZ corners. Figure~\ref{fig:fig3}a-b reproduce the measured dispersions near \mx~and \my, respectively. As temperature is raised, the \dyz~hole-like band at \mx~shifts down (Fig.~\ref{fig:fig3}c) as the \dxz~hole-like band at \my~shifts up (Fig.~\ref{fig:fig3}d), merging at $\sim$-30 meV at \ts, when \cfour~rotational symmetry is restored. The energy difference between these two bands as a function of temperature follows an order parameter-like behavior, with a full strength of 50 meV deep in the nematic state (Fig.~\ref{fig:fig3}g). In addition, the lower band at $\sim$~-50 meV at \mx~(Fig.~\ref{fig:fig3}c) also follows the behavior of the \dxz~band at \my~(Fig.~\ref{fig:fig3}d). We will discuss this later when we introduce the complete band reconstruction presented in the next section.

\begin{figure*}
\includegraphics[width=0.95\textwidth]{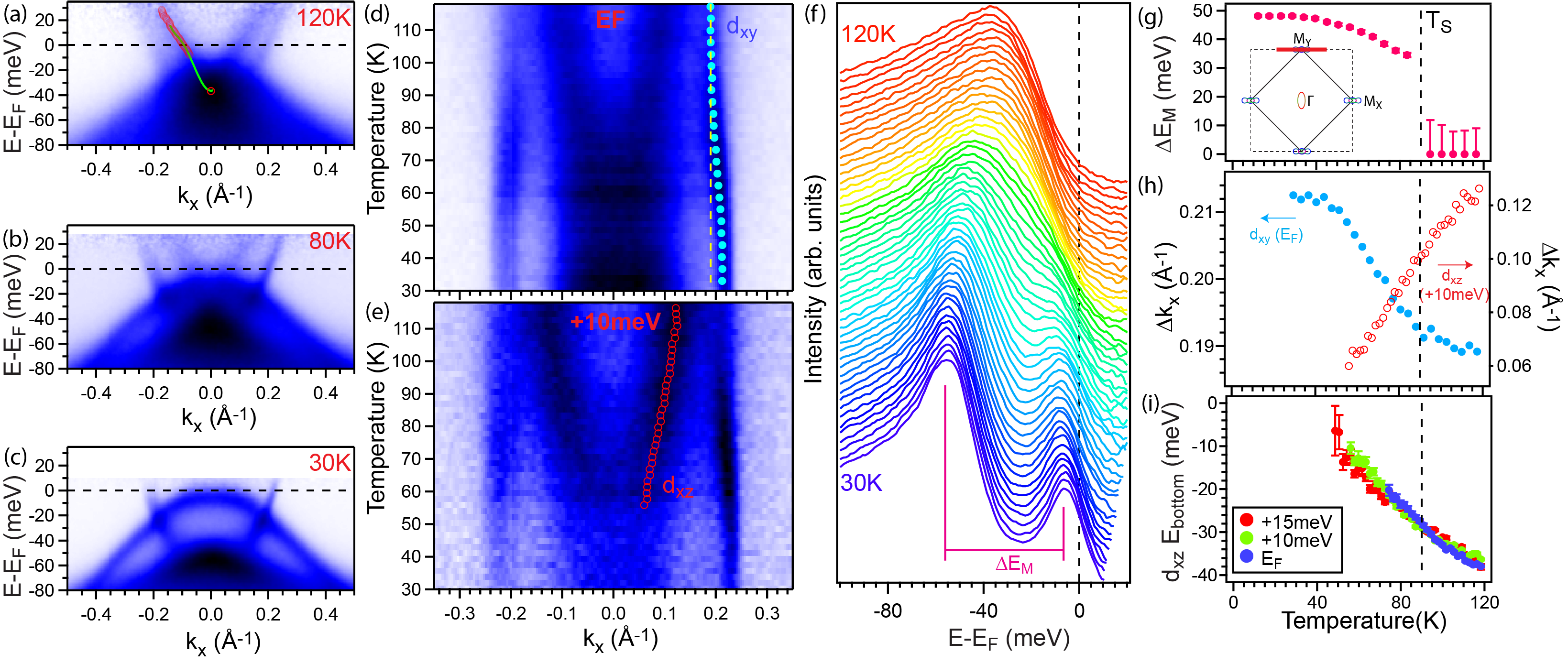}
\caption{\label{fig:fig_tdep} Temperature evolution of the electron bands orthogonal to \gmy. (a)-(c) Raw spectra taken at selected temperatures across \my~on a detwinned FeSe. The cut direction is shown in the inset of panel (g), perpendicular to the \gmy~high symmetry. Polarization is even with respect to the cut. All cuts have been divided by the Fermi-Dirac function convolved with the instrumental resolution. Fitted MDC peaks of the inner electron band are shown for the left half (red circles), along with the fitted even function for the inner electron band (green). (d) Temperature evolution of the MDC taken at \ef~of the cut in (a). Yellow dotted line marks the outer MDC peak position at 120 K for reference. Fitted MDC peaks for the outer \dxy~electron band are shown. (e) Temperature evolution of the MDC taken at +10 meV of the cut in (a), with fitted MDC peaks for the inner \dxz~band shown. (f) The EDCs taken at the \my~point in (a) taken from 30 K to 120 K. (g) Fitted peak separation in energy in (f) as a function of temperature. (h) Fitted MDC peak positions for the outer \dxy~and the inner \dxz~electron bands reproduced from (d)-(e). (i) The projected energy of the inner \dxz~electron band bottom with temperature estimated from the shift in MDC peaks and the k to E conversion based on the assumption of a rigid band shift from the 120 K data in (a).}
\end{figure*}

Next, we track the behavior of the \dxy~portion of the electron pocket near the \mx~point from a temperature-dependent measurement centered at \my~along the direction orthogonal to the \gmy~high symmetry direction (Fig.~\ref{fig:fig_tdep}). In this set of measurements, the two peaks in the EDCs from \dxz~and \dyz~at the \my~point again merge into one abruptly across a characteristic temperature,  90 K, benchmarking the nematic phase transition for this strained sample (Fig.~\ref{fig:fig_tdep}f,g). In the normal state at 120 K, two electron bands are clearly seen crossing \ef~as expected (Fig.~\ref{fig:fig_tdep}(a)), with the outer one being of \dxy~character and the inner one \dxz~character~\cite{Watson2015,Watson2015b,Watson2016c}. We first discuss the outer \dxy~electron band. By fitting the momentum distribution curve (MDC) at \ef~to obtain the band crossings, the \kf~points, we observe an expansion of this portion of the Fermi pocket with lowering temperature (Fig.~\ref{fig:fig_tdep}h). This expansion indicates that the \dxy~electron band shifts down in energy upon entering the nematic phase. As this measurement is perpendicular to the strained direction (longer a axis), the outer \dxy~electron band forms the tips of the peanut-shaped electron pocket that originates from the \mx~point of the 1-Fe BZ (Fig.~\ref{fig:fig4}b). The observed expansion is consistent with previous report~\cite{Watson2016c}, indicating the participation of the \dxy~orbital in the nematic order. Since \dxy~is a \cfour~symmetric orbital, anisotropy in \dxy~must appear via a hopping term, which causes the \dxy~states at \mx~and \my~to shift in opposite directions in energy~\cite{Su2015a,Fernandes2014e,Christensen2019}. The observation of the downward shift of the \dxy~originating from the \mx~point suggests that the shift at \my~is upward in energy.

Finally, we discuss the behavior of the \dxz~electron band. The temperature-evolution of the MDC taken at \ef~shows that the \kf~points of the inner \dxz~electron band move closer together as temperature is lowered (Fig.~\ref{fig:fig_tdep}d). This indicates that the inner electron band shifts up in energy, which can again be quantified via tracking of the MDC peaks. However, such analysis at \ef~is complicated by the contribution of the \dyz~hole-like band that also approaches \ef. To avoid such complication, we analyze the MDCs taken at 10 meV above \ef~after dividing the spectra by the Fermi-Dirac function convolved with the instrumental energy resolution (Fig.~\ref{fig:fig_tdep}e). The corresponding \dxz~MDC peak positions as a function of temperature (Fig.~\ref{fig:fig_tdep}h) show the shift of the inner electron band to be gradual through \ts, in stark contrast to the order-parameter-like abrupt splitting of the \dxz~and \dyz~states at M (Fig.~\ref{fig:fig3}g and Fig.~\ref{fig:fig_tdep}g). This observed gradual behavior may be due to a combination of effects, the specific detailed nature of which remains to be fully understood. In particular, the inner electron band bottom is still visible below \ef~across \ts, as can be seen in the data shown for 80 K (Fig.~\ref{fig:fig_tdep}b). To estimate the temperature at which the electron band is lifted to above \ef, we can convert the observed shift in momentum to shift in energy using the $E(k)$ dispersion relation fitted from this band at 120 K (Fig.~\ref{fig:fig_tdep}a) while assuming a rigid band shift. Such a $k$-to-$E$ conversion allows us to estimate the position of the electron band bottom from a fitting of the MDC peaks of the inner electron band at any energy. The results for MDC analyses done at \ef, +10 meV, and +15 meV are shown in Fig.~\ref{fig:fig_tdep}i. By extrapolation, the band bottom is estimated to cross \ef~below $\sim$30 K, which may be consistent with the suggested Lifshitz transition reported in a muon spin rotation experiment~\cite{Grinenko2018}. We caution that this rigid band estimate is a conservative lower bound of the shift in energy as the band bottom rises faster than the upper branch of the electron band due to a hybridization effect that will be discussed in the next section.

\subsection{Schematic of nematic reconstruction}
\begin{figure*}
\includegraphics[width=\textwidth]{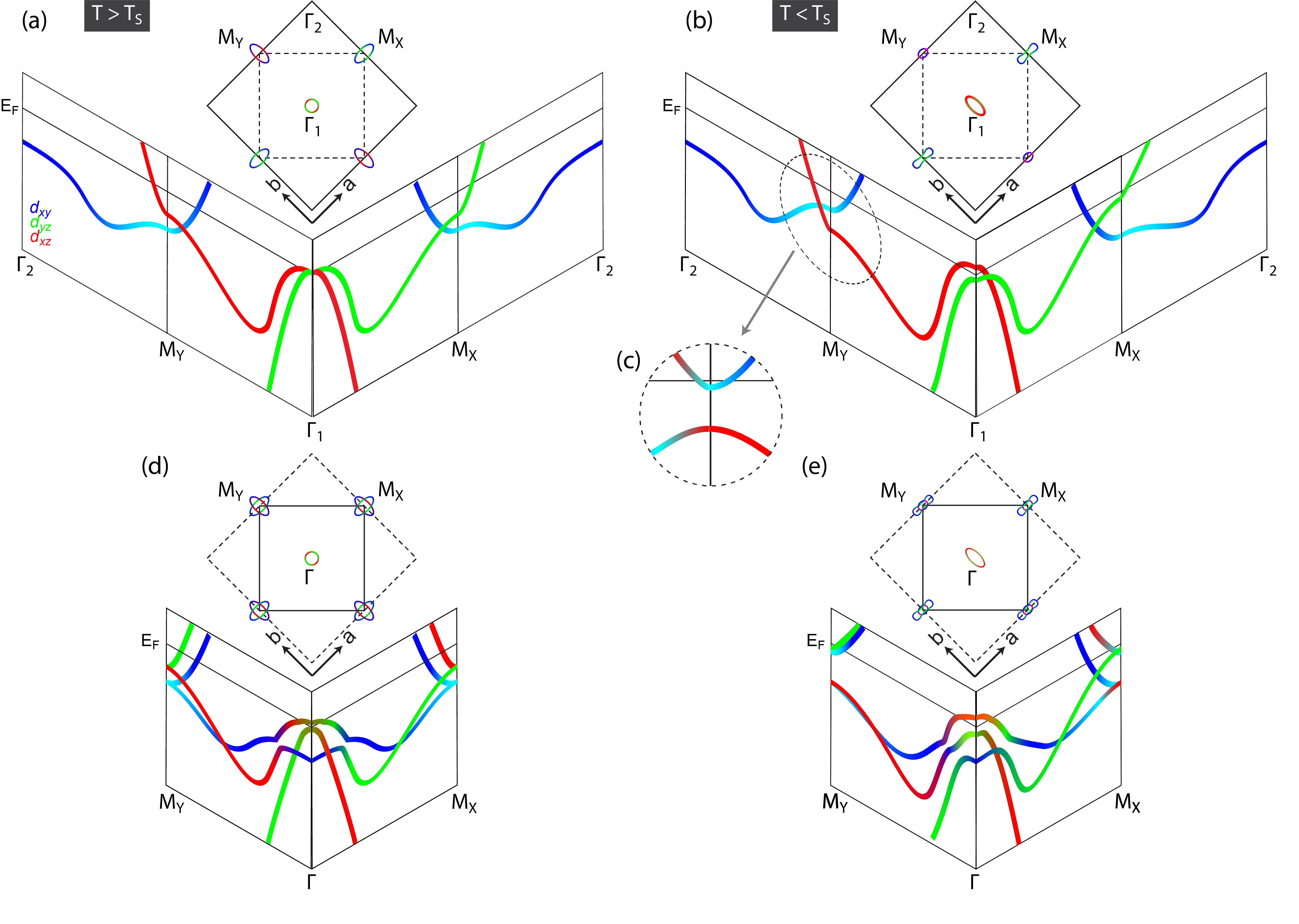}
\caption{\label{fig:fig4} Schematic of the nematic band reconstruction. (a)-(b) Summary of band structure in the tetragonal (a) and orthorhombic (b) phases in the unfolded 1-Fe BZ. Band hybridizations and the effect of SOC are omitted for simplicity. (c) Note that at \my~in the nematic phase, the downshifted \dxz~band crosses the \dxy~band, opening up a hybridization gap. (d)-(e) Summary of band structure in the tetragonal and orthorhombic phases in the folded 2-Fe BZ. Band hybridizations and SOC are incorporated here for direct comparison to data. The colors correspond to: \dxz~(red), \dyz~(green), \dxy~(blue, with cyan marking portions with weak photoemission matrix elements). }
\end{figure*}

Given the above three observed changes across \ts, the following three key aspects of the electronic reconstruction must occur in the nematic order in FeSe: i) momentum-dependent orbital anisotropy between \dxz~and \dyz; ii) direct involvement of the \dxy~orbital via an anisotropic hopping term; and iii) shrinking of the electron pocket containing \dxz~that originates from the \my~point of the 1-Fe BZ. 

Figure~\ref{fig:fig4} presents a schematic of the changes to the electronic structure that is compatible with all of these observations. This is not a calculation but our best understanding of the nematic band reconstruction based on the data. For simplicity, we first discuss the band reconstruction in the unfolded 1-Fe BZ without considerations of band hybridizations and SOC. In the tetragonal state (Fig.~\ref{fig:fig4}a), \cfour~symmetry is respected, seen in the degeneracy of the \dxz~and \dyz~orbitals along the orthogonal directions. We note that the \dxy~states (band bottom of the electron band and the band top of the hole band) near the M points of the BZ are always observed to have weak photoemisson matrix elements across iron-based superconductor families. Specifically, intensity of the electron-like \dxy~band at M is often observed to disappear as the band bottom is approached, as does the \dxy~hole band from the \g~point as it approaches the M point~\cite{Brouet2012}. This is especially true for materials where the \dxy~electron band bottom is well below that of the \dxz/\dyz~band. Hence this portion of the \dxy~band appears to have different matrix elements than the observable portion of \dxy, and we shade it light blue for this discussion.

Below \ts~(Fig.~\ref{fig:fig4}b), the degeneracy between \dxz~and \dyz~bands is lifted, but in a momentum-dependent fashion. Near \g, the \dxz~hole band top shifts up while the \dyz~hole band top shifts down to below \ef. This anisotropy between \dxz~and \dyz~is reversed at \mx~and \my~with a much bigger magnitude. This k-dependent reversal of orbital anisotropy is consistent with previous reports~\cite{Suzuki2015a,Zhang2015a,Zhang2016c,Fanfarillo2016a,Pfau2019}. When incorporating the additional effects of the SOC and band hybridizations in the complete folded 2-Fe BZ (Fig.~\ref{fig:fig3}d-e), the bands near \g~acquire mixed orbital characters where they cross and hybridize. However, the incorporation of the SOC does not modify the \cfour~or \ctwo~symmetry of the electronic structure in the tetragonal or orthorhombic states, respectively.

Next, we focus our discussion to the continuous \dxz~band along the $\Gamma_1-M_Y-\Gamma_2$ path in Fig.~\ref{fig:fig4}b. As we have directly observed the downward shift of the \dxz~hole band along $\Gamma_1-M_Y$ as well as the upward shift of the \dxz~electron band along $M_Y-\Gamma_2$, we come to the conclusion that the key to reconcile such apparent contradictory shift of a continuous band within a small momentum range is that the band is no longer continuous in the nematic state. This can be naturally explained via a hybridization between the \dxz~band and the \dxy~band which inverts in energy when the \dxz~band shifts down in energy with the onset of the nematic order. The anisotropic hopping nematicity in \dxy~would also shift the \dxy~band up at the \my~point, further contributing to this band inversion. As a result, these two bands cross along \mygtwo~in the nematic state (compare dotted circle in Fig.~\ref{fig:fig4}b with Fig.~\ref{fig:fig4}a). The \dxz~and \dxy~bands have opposite parity along \gmy. Therefore, they do not hybridize at the crossing along this direction in the normal state. However, these bands have the same parity along \mygtwo~and hence would hybridize at the crossing point in the nematic state~\cite{Lee2008,Brouet2012,Nica2015b}. As a result, the original \dxz~electron band and \dxy~hole band at \my~swap characters such that near \my~the hole-like band acquires \dxz~character and the bottom of the \dxz~electron band acquires \dxy~character. Due to the weak matrix elements of the \dxy~band near the zone corner (marked as light blue) as discussed previously, the band bottom of the inner electron band becomes weaker in intensity in the nematic phase, which may give rise to the impression of incoherence of the \dxz~orbital in the nematic state. On the other hand, the hole-like band now acquires \dxz~character from the original electron band, the photoemisson matrix elements of which under parity switching~\cite{Brouet2012} allow it to be observed simultaneously as the \dyz~hole-like band, which is consistent with the intensity pattern of the lower hole band shown in Fig.~\ref{fig:fig2}b.

The exchange of orbital character due to this hybridization is the origin of the peculiar temperature dependence we observe at \mx~in Fig.~\ref{fig:fig3}c,e. Since the measurement always shows the 2-Fe folded BZ, we compared our data to that of the schematic in Fig.~\ref{fig:fig4}e. The hole band at -50meV near \mx~contains considerable spectral weight from the \dxz~electron band and therefore follows the same temperature dependence as the \dxz~band at \my~in Fig.~\ref{fig:fig3}d,f. Note that this band inversion does not occur between the \dyz~and \dxy~band along \mxgtwo~since the \dyz~and \dxy~bands there moves apart in energy and the two bands never cross. Also, a gap does not open at the crossing between \dyz~and \dxy~along \gmx~since \dyz~and \dxy~have opposite parities along \gmx~under the glide mirror symmetry~\cite{Lee2008,Brouet2012,Nica2015b}, forming the reported Dirac cones~\cite{Tan2016}. We would like to point out that the gap opening due to inversion between \dxz~and \dxy~is purely a band hybridization effect as a result of nematicity-driven band shifting~\cite{Hao2014d}. Similar behavior is observed in more strongly correlated iron chalcogenides close to the orbital-selective Mott phase~\cite{Yi2013,Yi2015a}, where stronger renormalization of the \dxy~orbital compared to that of \dxz~and \dyz~also inverts the energy positions of the \dxy~hole band and \dxz/\dyz~electron bands at the BZ corner, opening up a similar hybridization gap in monolayer FeSe/\sto~film, bulk Fe(Te,Se), and \AFS, albeit driven by correlations rather than nematicity.

\subsection{Hybridization between \dxz~and \dxy~bands}

\begin{figure}
\includegraphics[width=0.45\textwidth]{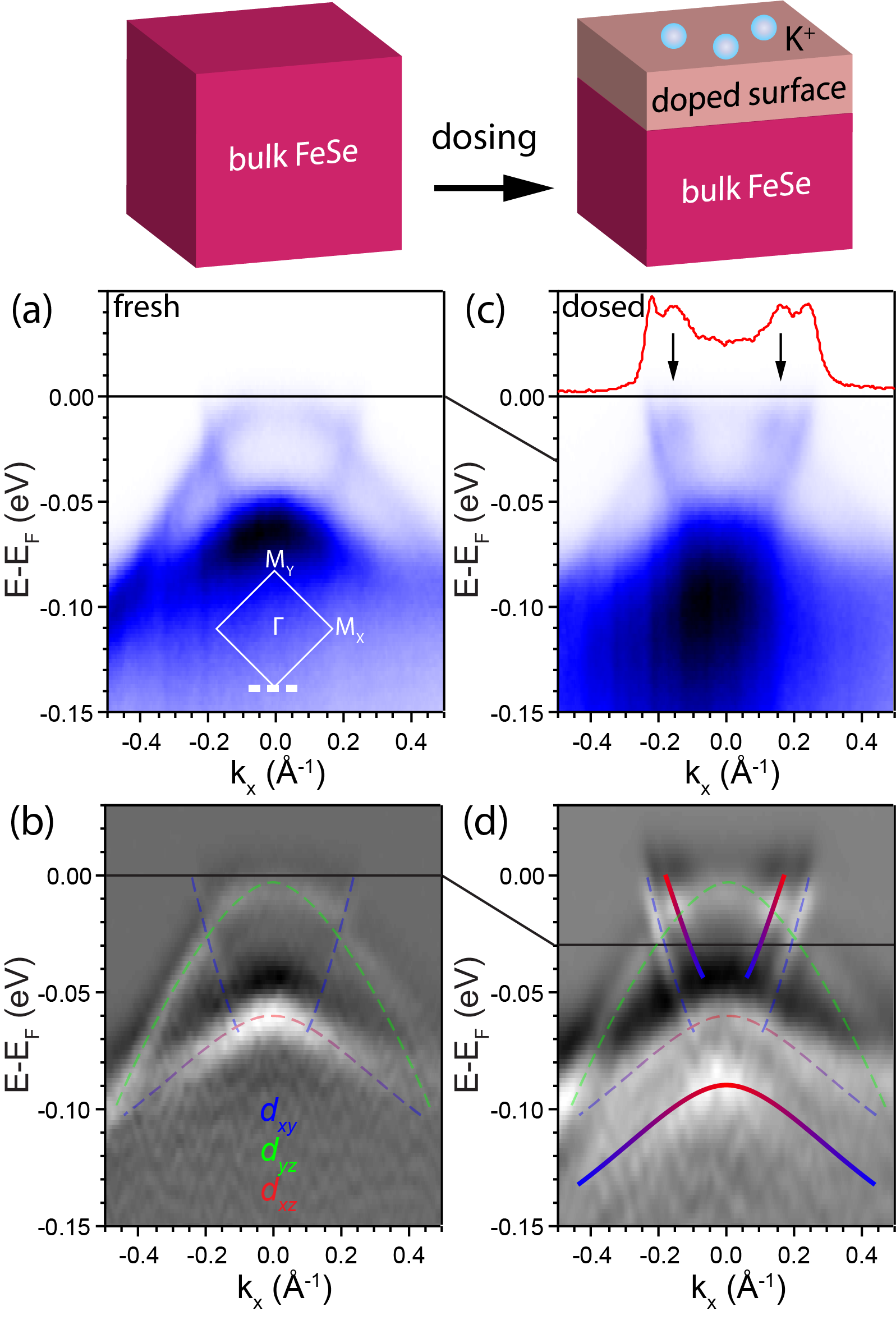}
\caption{\label{fig:fig6} Finding the missing second electron pocket by surface-doping. (a) Raw spectra taken across \my~on a freshly cleaved detwinned FeSe, at 56 eV. (b) Second energy derivative of (a) with schematic of band dispersions. (c) Same measurement as (a) but after doping the surface with potassium. The MDC at \ef~is plotted in red, with arrows indicating the new electron band induced by potassium doping. (d) Second energy derivative of (c) where the bands from the undoped bulk from (b) are reproduced as dotted lines and the new doped surface bands are marked by solid lines.}
\end{figure}

To further confirm the hybridization picture of the \dxz~and \dxy~band at \my, which pushes the part of the electron band with \dxz~orbital character above \ef~at low temperature, we surface-dope completely detwinned FeSe via deposition of K atoms. Due to the orbital character switching, the electron band at \my~has a band bottom with \dxy~character with weak photoemission matrix elements. Based on the temperature dependence data shown in Fig.~\ref{fig:fig_tdep}, the \dxz~portion of the electron band is pushed to above \ef~at low temperature, which should be detectable with sufficient electron doping (Fig.~\ref{fig:fig6}). As has been reported, the charge carriers added in FeSe due to surface-doping are strongly localized to the top surface layer~\cite{Wen2016a,Seo2016,Ye2015}. In the photoemission process, the surface-doped sample would exhibit two sets of bands: one set from the undoped bulk and another set from the electron-doped top surface. Figure~\ref{fig:fig6}a shows the measured dispersions across the \my~point on a freshly cleaved sample. After K doping, the same measurement shows that in addition to the unshifted bulk band structure (dotted lines in Fig.~\ref{fig:fig6}d), a new set of bands emerged that originates from the doped surface (solid lines in Fig.~\ref{fig:fig6}d). Compared with bulk bands, the lower hole-like band has shifted down in energy by 30 meV while a new electron band has appeared. This is the missing electron band. Notably, this new electron band has an intensity profile that is very weak at the band bottom and becomes stronger approaching \ef~(Fig.~\ref{fig:fig6}c-d), fully consistent with the understanding of the changing orbital character from \dxz~to \dxy~near the band bottom, confirming the mechanism of hybridization between the inverted \dxz~and \dxy~bands at \my~(Fig.~\ref{fig:fig4}b).

\section{Discussions and Summary}

Based on all the observations together, we come to the following main conclusions regarding the nematicity in FeSe:
\begin{itemize}
	\item The nematic energy scale between \dxz~and \dyz~is momentum-dependent and reaches $\sim$50 meV at the BZ corner.
	\item The \dxy orbital shows a nematic band shift and hence participates in the nematic order. This behavior can be explained by an anisotropic hopping term such that \dxy~shifts down in energy at \mx~and shifts up at \my~in the 1-Fe BZ representation.
	\item The electron pocket at \my~containing \dxy~and \dxz~orbital in the tetragonal state shrinks as temperature is lowered across \ts. This can be explained by a hybridization between the \dxy~and \dxz~bands near \my~from the nematic band shifts of the associated orbitals.
\end{itemize}

The confirmation of the magnitude of the nematic energy scale between \dxz~and \dyz~implies that the nematic order in FeSe without the presence of static magnetic order is similar to that in iron-pnictides where the nematic order and magnetic order are strongly coupled. This is consistent also with a recent study where the form of momentum-dependence of the nematic band shift is shown to be comparable between FeSe and \bfa~\cite{Pfau2019}. We emphasize that correct identification of the nematic energy scale presented here hinges on the correct identification of the orbital characters of the observed bands, for which complete detwinning of the crystal is crucial. Furthermore, a detailed temperature dependence data of the \mygtwo~cut with the even parity polarization on a completely detwinned crystal is the key to detect the electron band with \dxz~orbital character, which is further confirmed by the electron doping with K deposition.

The participation of the \dxy~orbital in the nematic order was previously reported for muti-layer FeSe film~\cite{Zhang2016c} and bulk FeSe~\cite{Watson2016c}. The anisotropic hopping for \dxy, and more generally for \dxz~and \dyz, has been discussed theoretically~\cite{Su2015a,Fernandes2014e,Christensen2019}. In addition, calculations including fluctuations beyond random phase approximation method in the multiorbital Hubbard model have shown that \dxy~dominates the contribution to the nematic susceptibility~\cite{Christensen2016}. Here we also show that the participation of the \dxy~orbital in nematicity is important in the observed shrinking and decreasing in the \dxz~spectral weight of the electron pocket at \my~via the hybridization effect.

The ``missing" electron pocket at \my~has been heavily discussed in the literature and forms the basis of a number of theoretical proposals for FeSe. It has been shown that the largest pairing interaction at low energies is between the hole and electron pockets across the BZ~\cite{Graser2009,Wang2009a,Chubukov2012}. Therefore when one of these electron pockets, in particular the one carrying \dxz~spectral weight from the \my~point of the 1-Fe BZ, has a drastically reduced presence at \ef, the pairing interactions are also much affected, leading to a very anisotropic and orbital-dependent pairing gap. Thus the missing electron pocket has acted as a key to many theoretical proposals on superconductivity in FeSe in order to explain the strongly anisotropic pairing gap, whether in the form of suppressed \dxz~spectral weight~\cite{Sprau2017,Hu2018,Yu2018,Kreisel2019} or missing channel for scattering between \gmy~\cite{Kreisel2015,Kang2018,Benfatto2018,Fanfarillo2018}. The cause of it had been mysterious.

The observations we have presented here confirm the dominance of the \dyz~electron pocket and the much smaller or even non-existent \dxz~electron pocket in the nematic phase. Importantly, we clearly identified the cause of this occurrence to be a band hybridization effect from a \dxz/\dxy~band inversion at \my~directly caused by an orbital-dependent band shift in the nematic phase. In addition, the \dxz~electron band is clearly visible down to at least 70K. From these observations, it seems unnecessay and unlikely that the \dxz~electron band becomes strongly incoherent. Furthermore, if we compare the measured bandwidth of the \dxz~and \dyz~bands in the nematic phase (Fig.~\ref{fig:fig2}a,e) with Density Functional Theory (DFT)~\cite{Subedi2008}, the bandwidth renormalization factor for both orbitals are around 4, giving a naive estimation of the coherence factor Z$\sim$0.25 for both \dxz~and \dyz. However, we do not exclude the possibility that in an energy scale very close to \ef~the Fermi velocities for \dxz~and \dyz~are different. For the \dxy~orbital, the band slope of the hole-like component between \g~and M measured in the nematic phase compared to DFT band structure calculated for the tetragonal state is renormalized by a factor of 8.9 along \gmx~and 6.7 along \gmy, and 7.4 when comparing the measured dispersion and calculated dispersion in the tetragonal phase, giving an estimation of the Z factor for \dxy~of 0.14, roughly half of that of the \dxz/\dyz~orbitals.

Interestingly, as a result of the k-dependent and orbital-dependent band shift caused by nematicity, the spectral weight of \dxz~and \dyz~are redistributed across the BZ such that \dxz~has a suppressed presence at \my~while \dyz~maintains presence at both \g~and \mx~(Fig.~\ref{fig:fig4}). For the \dyz~orbital near the \g~point, even though there is very little spectral weight at \ef, \dyz~density of states is quickly recovered below \ef~by the contribution of the hole band that appears immediately below \ef. At \mx, the \dyz~electron band provides finite spectral weight of \dyz~within an energy scale of the bandwidth about \ef. Therefore deep in the nematic state, there remains sufficient electronic states of \dyz~orbital for scattering between \g~and \mx. For the \dxz~orbital, while its spectral weight dominates at \g, the band inversion and hybridization at \my~caused by the nematic band shift produces a large effective gap in the density of states from \dxz~(Fig.~\ref{fig:fig4}c), depleting the available states for the electrons to be scattered from \g. Effectively, the disappearance of the \my~electron pocket in the superconducting state deep in the nematic phase turns off the scattering between \g~and \my, which could be consistent with an inequivalent intra-orbital scattering between \dxz~and \dyz~orbitals, or selective-scattering only between \gmx. This is consistent with recent inelastic neutron scattering measurement revealing strongly anisotropic low energy magnetic excitations and spin resonance selectively appearing at ($\pi$,0)~\cite{Chen2019}. This suppressed \gmy~scattering in the \dxz~orbital due to the combination of nematic band shift and hybridization effect could be an alternative mechanism for the manifested orbital-selective Cooper pairing~\cite{Sprau2017}. 

In summary, we have presented comprehensive temperature-dependent study of the electronic structure of detwinned FeSe across the nematic phase transition. We have clarified the orbital characters of all of the bands and identified the nematic energy scale at the BZ corner to be 50 meV, which is a substantial portion of the renormalized bandwidth. In addition, we clearly observed the disappearance of the "missing" electron pocket through the nematic phase transition via an upshift in energy. This observation is consistent with all of our band assignment and temperature-dependent behavior. We have identified the cause of the disappearance of the \dxz~electron pocket to be the nematicity-induced band inversion between the \dxz~and \dxy~bands at \my. This nematic reconstruction of the low energy band structure causes a dramatic redistribution of the \dxz~and \dyz~spectral weight across the BZ. As a result, such rearrangements of orbital-dependent electronic states may strongly modify the intra-orbital scattering across the BZ, providing the basis for the strongly anisotropic pairing states observed in FeSe. 

\section{acknowledgments}
We thank Timur Kim, Luke Rhodes, Matthew Watson, Amalia Coldea, Morten Christensen, Andrey Chubukov, Rafael Fernandes, Roser Valenti, Andreas Kreisel, Brian Andersen, Peter Hirschfeld, and Laura Fanfarillo for congenial discussions. ARPES experiments were performed at the Stanford Synchrotron Radiation Lightsource, which is operated by the Office of Basic Energy Sciences, U.S. DOE. Work at University of California, Berkeley and Lawrence Berkeley National Laboratory was funded by the U.S. Department of Energy, Office of Science, Office of Basic Energy Sciences, Materials Sciences and Engineering Division under Contract No. DE-AC02-05-CH11231 within the Quantum Materials Program (KC2202) and the Office of Basic Energy Sciences. The ARPES work at Rice University was supported by the Robert A. Welch Foundation Grant No. C-2024 (M.Y.) as well as the Alfred P. Sloan Foundation. The work at Stanford Institute of Materials and Energy Sciences is supported by the DOE Office of Basic Energy Sciences, Division of Materials Sciences. The FeSe single crystal growth work at Rice University is supported by the U.S. DOE, BES under Contract No. DE-SC0012311 (P.D.). A part of the material characterization work at Rice University is supported by the Robert A. Welch Foundation Grant No. C-1839 (P.D.). Theory work at Rice University is supported by the U.S. Department of Energy, Office of Science, Basic Energy Sciences, under Award No. DE-SC0018197, and by the Robert A. Welch Foundation Grant No. C-1411. Work at Renmin University is supported by the National Science Foundation of China Grant numbers 11374361 and 11674392 and Ministry of Science and Technology of China, National Program on Key Research Project Grant number 2016YFA0300504. H.P. acknowledges support from the German Science Foundation (DFG) under reference PF 947/1-1. Y. H. acknowledges support from the Miller Institute for Basic Research in Science.

%

\end{document}